\documentclass{desyproc}

\begin{document}
%------------------------------------
\title{Searching for tetraquarks on the lattice}

%for single authors the superscripts are optional
\author{{\slshape S. Prelovsek$^1$, T. Draper$^2$, C.B. Lang$^3$, M. Limmer$^3$, K.-F. Liu$^2$, N. Mathur$^4$ and D. Mohler$^5$ }\\[1ex]
$^1$ Department of Physics, University of Ljubljana and Jozef Stefan Institute,    Slovenia. \\
$^2$ Department of Physics and Astronomy, University of Kentucky, Lexington, KY 40506, USA.\\
$^3$ Institut f\"{u}r Physik, FB Theoretische Physik, Universit\"{a}t Graz, A-8010 Graz, Austria.\\
$^4$ Department of Theoretical Physics, Tata Institute of Fundamental Research, Mumbai, India.\\
$^5$ TRIUMF, 4004 Wesbrook Mall Vancouver, BC V6T 2A3 Canada, Canada}

% please enter the contribution ID for the DOI
\contribID{xy}

% TO THE CONFERENCE EDITORS: 
% please update the following information      
% before sending the template to the authors
\confID{800}  % if the conference is on Indico uncomment this line
\desyproc{DESY-PROC-2009-xx}
\acronym{LP09} % if you want the Acronym in the page footer uncomment this line
\doi  % if there is an online version we will register DOIs

\maketitle

\begin{abstract}
We address the question whether the lightest scalar mesons $\sigma$ and  $\kappa$ are  tetraquarks.  We present a search for possible light tetraquark states with $J^{PC}=0^{++}$ and $I=0,~1/2,~3/2,~2$ in the dynamical and the quenched lattice simulations using tetraquark interpolators.   In all the channels, we unavoidably find lowest  scattering states $\pi(k)\pi(-k)$ or $K(k)\pi(-k)$ with back-to-back momentum $k=0,~2\pi/L,\cdots$. However, we find an additional light state in the $I=0$ and $I=1/2$ channels, which may be related to the observed resonances $\sigma$ and $\kappa$ with a strong tetraquark component.  In the exotic repulsive channels $I=2$ and $I=3/2$, where no resonance is observed, we find no light state in addition to the scattering  states. 
\end{abstract}

It is still not known whether the lightest observed nonet of scalar mesons 
$\sigma$, $\kappa$,  $a_0(980)$ and $f_0(980)$ \cite{pdg} are conventional $\bar qq$ states or exotic tetraquark $\bar q\bar qqq$ states. Tetraquark interpretation was proposed by  Jaffe back in 1977 \cite{jaffe77} and it is supported by many 
phenomenological studies, for example \cite{pdg,phenomenology}. The tetraquarks, composed of a scalar diquark  and anti-diquark, form a flavor nonet and are expected to be light. 
The observed ordering $m_\kappa<m_{a_0(980)}$ favors  tetraquark interpretation since  the $I=1$ state $[\bar d\bar s][us]$ with additional valence pair $\bar ss$ is naturally heavier than the $I=1/2$ state $[\bar s\bar d][du]$. 

It is   important to determine whether QCD predicts any scalar tetraquark states below $1$ GeV from a first principle lattice QCD calculation. Previous lattice simulations \cite{mathur_scalar,tetra_lat} have not given the final answer  yet. The  strongest claim for $\sigma$ as tetraquark was obtained   using the sequential Bayes method to extract the spectrum \cite{mathur_scalar} and needs confirmation using a different method.  Our new results, given in this proceeding, are presented with more details in \cite{tetra_dyn_lat09,tetra_dyn_paper}. 

We calculate the energy spectrum of scalar tetraquark states with $I=0,~2, ~1/2, ~3/2$ in dynamical and quenched lattice simulations.   
Our dynamical simulation ($a\!\simeq\! 0.15$ fm, $V\!=\!16^3\!\times\! 32$) uses  dynamical Chirally Improved $u/d$ quarks \cite{ci_dyn} and it is the first dynamical simulation intended to study tetraquarks.  The quenched simulation ($a\!\simeq \!0.20$ fm, $V\!=\! 16^3\!\times\! 28$)   uses overlap fermions, which have exact chiral symmetry even at finite $a$. 

The energies of the lowest three physical states are extracted from the correlation functions 
$C_{ij}(t)=\langle 0| {\cal O}_i(t){\cal O}^{\dagger}_j(0)|0\rangle_{\vec p =\vec 0}= \sum_n Z_i^{n}Z_j^{n*}e^{-E_n~t}$ with tetraquark interpolators ${\cal O}\sim \bar q\bar qqq$, where $Z_i^n\equiv \langle 0|{\cal O}_i|n\rangle$. In all the channels we use three different interpolators that are products of two color-singlet currents \cite{tetra_dyn_lat09}. In addition, we use  two types of diquark anti-diquark interpolators in $I=0,~1/2$ channels \cite{tetra_dyn_lat09}. 

When  calculating the $I=0,~1/2$ correlation matrix,  we neglect the so-called single and double disconnected quark contractions \cite{tetra_lat}, as in all previous tetraquark studies. The resulting  states have only a $\bar q\bar qqq$ Fock component in this approximation, while they would  contain also a $\bar qq$  component if single disconnected contractions  were taken into account \cite{tetra_lat}. Since we are searching for ``pure'' tetraquark states in this pioneering study, our approximation is  physically motivated.

All physical states $n$ with given $J^{PC}=0^{++}$ and $I$ propagate between the source and the sink in the correlation functions. Besides possible tetraquark states, there are unavoidable contributions from   scattering states $\pi(k)\pi(-k)$ for $I=0,~2$ and scattering states  $\pi(k)K(-k)$ for $I=1/2,~3/2$. Scattering states have discrete momenta  $\vec k=\tfrac{2\pi}{L}\vec j$ on the lattice of size $L$ and energy $(m_\pi^2+\vec k^2)^{1/2}+(m_{\pi,K}^2+\vec k^2)^{1/2}$ in the non-interacting approximation.   Our main question is whether we find some  light state in addition the scattering states in $I=0,~1/2$ channels. If such a state is found, it  could be related to the resonances $\sigma$ or $\kappa$ with a strong tetraquark component. 

The energies $E_n$ are extracted from the correlation functions $C_{ij}(t)$ via the  eigenvalues $\lambda^n(t)\propto e^{-E_n(t-t_0)}$ of the generalized eigenvalue problem  $C(t)\vec u^{n}(t)=\lambda^{n}(t,t_0) C(t_0)\vec u^{n}$  at some reference time $t_0$ \cite{var}. 

\begin{figure}[hb]
\begin{center}
\includegraphics[height=4.8cm,clip]{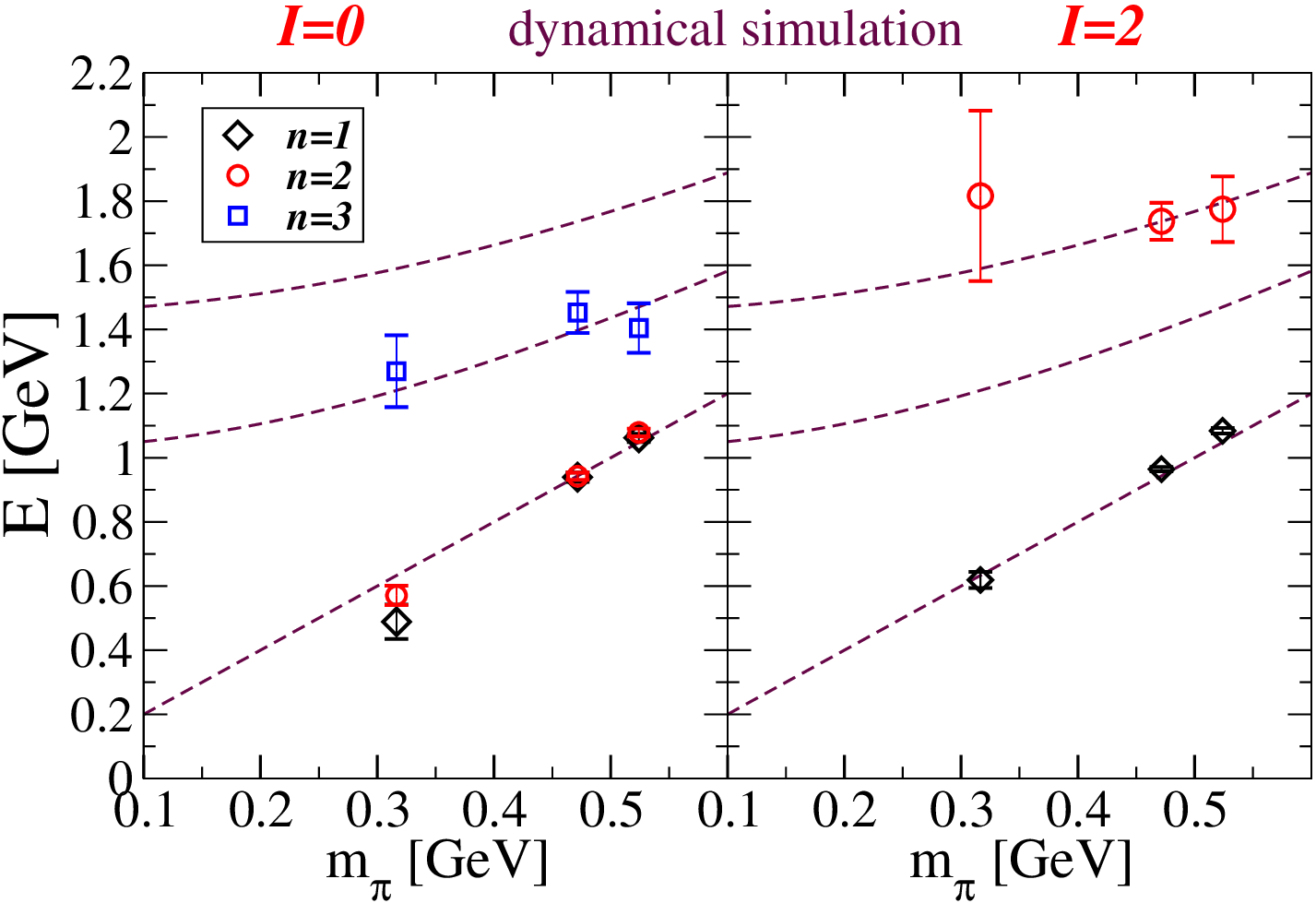}
\includegraphics[height=4.8cm,clip]{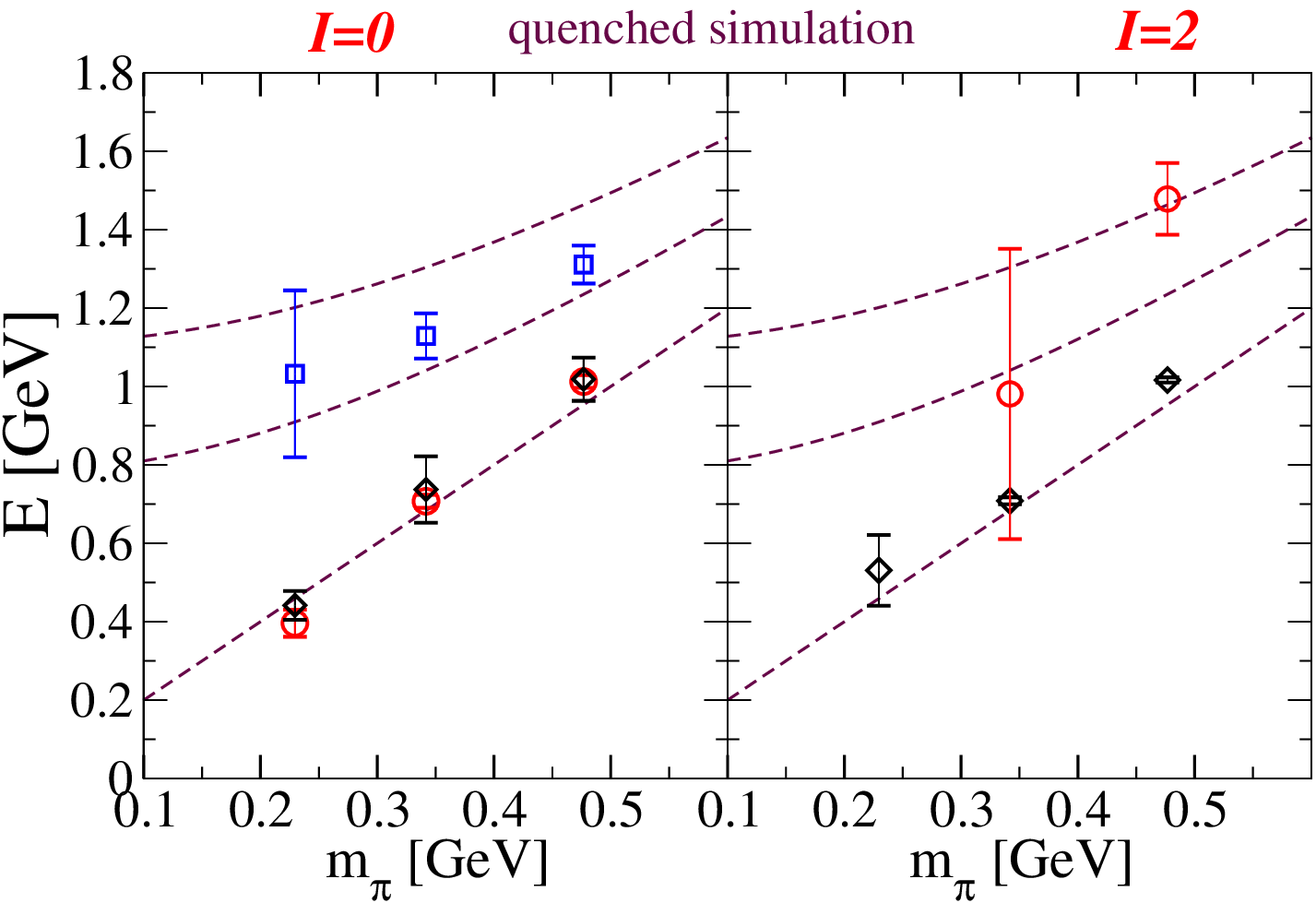}

$~$

\includegraphics[height=4.8cm,clip]{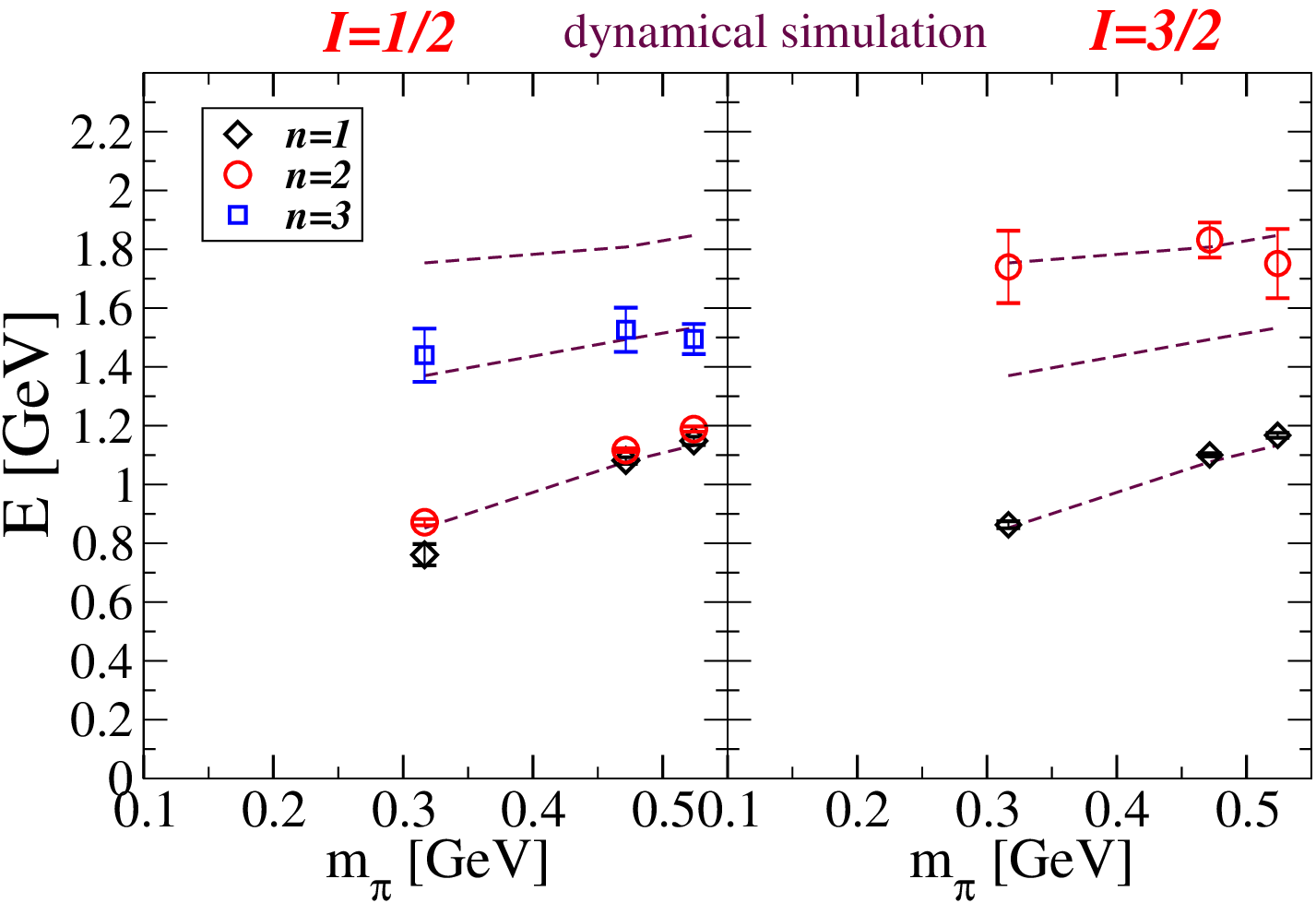}
\includegraphics[height=4.8cm,clip]{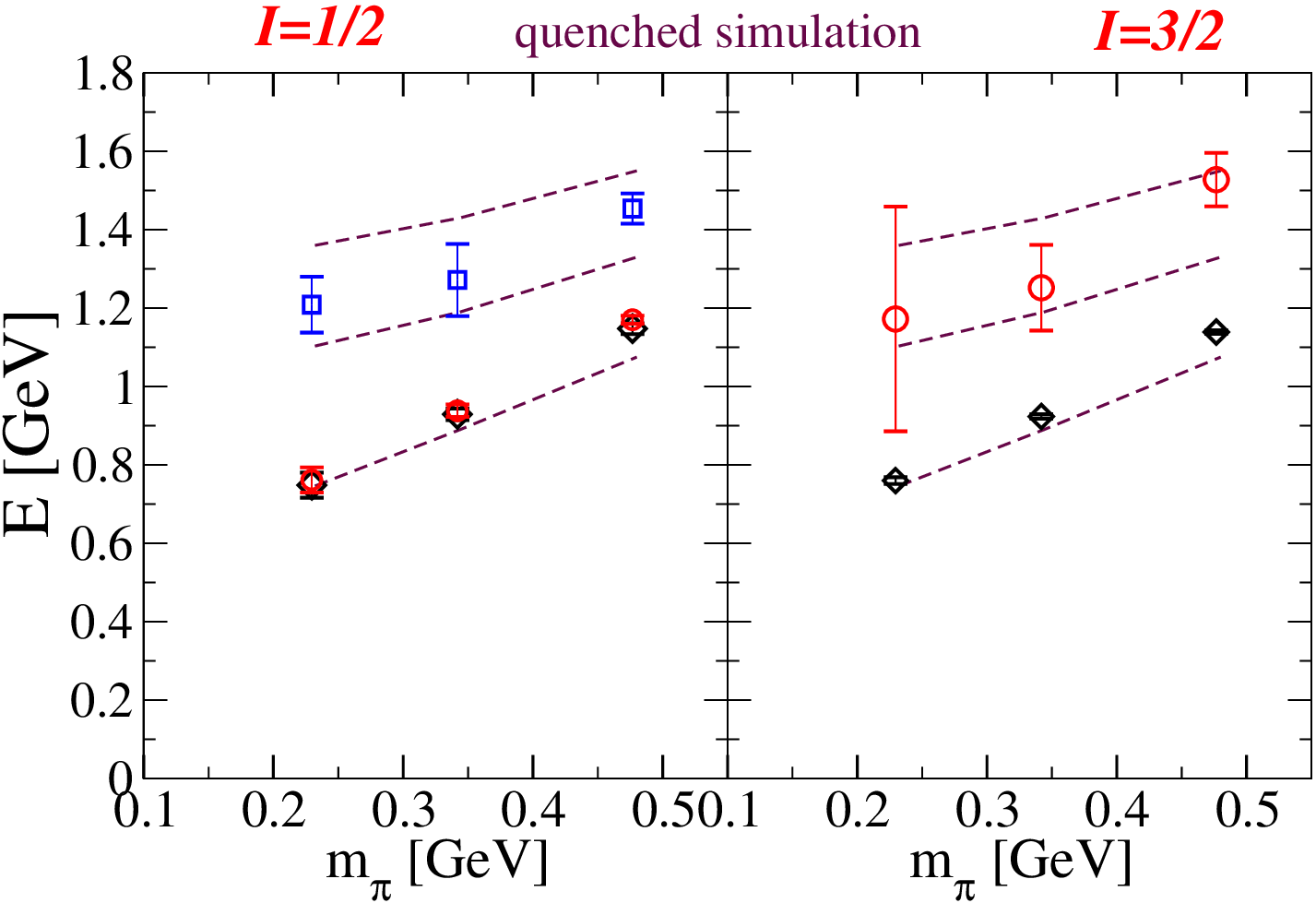}
\end{center}
\caption{ \small The resulting spectrum $E_{n}$ for $I=0,~2,~1/2,~3/2$ in the dynamical (left) and the quenched (right) simulations. Note that there are two states (black and red) close to each other in $I=0$ and $I=1/2$  cases. The lines at $I=0,~2$ present the energies of non-interacting $\pi(k)\pi(-k)$ with $k=j\tfrac{2\pi}{L}$ and $j=0,1,\sqrt{2}$. Similarly, lines at $I=1/2,~3/2$ present  energies of  $\pi(k)K(-k)$. }\label{fig_spect}
\end{figure}  
 
The resulting spectrum $E_n$  for all four isospins is shown in Fig. \ref{fig_spect}. The lines present the energies of the scattering states in the non-interacting approximation. Our dynamical and quenched results are in qualitative agreement. 

In the repulsive channel $I=2$, where no resonance is expected, we indeed find only the candidates for the scattering states $\pi(0)\pi(0)$ and $\pi(\tfrac{2\pi}{L}) \pi(-\tfrac{2\pi}{L})$ with no additional light state. The first excited state is higher than expected due to the smallness of $3\times 3$ basis.  Similar conclusion applies for the repulsive $I=3/2$ channel with $\pi K$ scattering states.

 In the attractive channel  $I=0$ we find two (orthogonal) states close to the threshold $2m_\pi$ and another state consistent with $\pi(\tfrac{2\pi}{L}) \pi(-\tfrac{2\pi}{L})$, so we do find an additional light state.  
This leads to a possible interpretation  that one of the two light states is the scattering state $\pi(0)\pi(0)$ and the other one corresponds to $\sigma$ resonance with strong tetraquark component (see more general discussion in \cite{sasaki}).   In the attractive $I=1/2$ channel we similarly find the candidates for the lowest two $\pi(k) K(-k)$ scattering states and a candidate for a $\kappa$ resonance with a large tetraquark component. These results have to be confirmed by another independent lattice simulation before making firm conclusions.

We investigate two criterion for distinguishing the one-particle (tetraquark) and two-particle (scattering) states  in \cite{tetra_dyn_paper}. The first criteria is related to the time dependence of $C_{ij}(t)$ and $\lambda^n(t)$  at finite temporal extent of the lattice. The second is related to the volume dependence of the couplings $\langle 0|{\cal O}_i|n\rangle$.    

The ultimate method to study $\sigma$ and $\kappa$ on the lattice would involve the study of the spectrum and couplings in presence of $\bar q\bar qqq\leftrightarrow \bar qq\leftrightarrow vac\leftrightarrow glue$ mixing, using interpolators that cover     these  Fock components. Such a study has to be done as a function of lattice size $L$ in order to extract the resonance  mass and width  using the L\"{u}scher's finite volume method \cite{sasaki,vol_dep}.  \\

{\bf Acknowledgments}

\vspace{0.1cm}

This work is supported by the Slovenian Research Agency, by the European RTN network FLAVIAnet (contract  MRTN-CT-035482), by the Slovenian-Austrian bilateral project (contract   BI-AT/09-10-012), by the USA DOE Grant DE-FG05-84ER40154, by the Austrian grant FWF DK W1203-N08 and by Natural Sciences and Engineering Research Council of Canada.

\begin{footnotesize}

\end{footnotesize}

\end{document}